# Chiral limit of QCD

Rajan Gupta [a]

[a]T-8 Group, MS B285, Los Alamos National Laboratory, Los Alamos, New Mexico 87545 U. S. A.

This talk contains an analysis of quenched chiral perturbation theory and its consequences. The chiral behavior of a number of quantities such as the pion mass $m_\pi^2$, the Bernard-Golterman ratios $R$ and $\chi$, the masses of nucleons, and the kaon B-parameter are examined to see if the singular terms induced by the additional Goldstone boson, $\eta'$, are visible in present data. The overall conclusion (different from what I presented at the lattice meeting) of this analysis is that, with some caveats on the extra terms induced by $\eta'$ loops, the standard expressions break down when extrapolating the quenched data with $m_q < m_s/2$ to physical light quarks. I then show that due to the single and double poles in the quenched $\eta'$, the axial charge of the proton cannot be calculated using the Adler-Bell-Jackiw anomaly condition. I conclude with a review of the status of the calculation of light quark masses from lattice QCD.

## 1. INTRODUCTION

The main question this review attempts to answer is "should the ostrich care about the alarmists view of quenched QCD"? The alarmists are two groups, Sharpe, Labrenz, and Zhang [3] [16] [18] and Bernard and Golterman [1] [2]. They have calculated, using quenched chiral perturbation theory, a number of quantities to 1-loop and find that in the quenched approximation $\eta'$ loops give rise to unphysical terms in the chiral expansion and that in many cases the chiral limit is singular. Also, the coefficients in the chiral expansion (including those of the normal chiral logs) are different in the full and quenched theories. The ostrich is everyone who wishes to continue using the chiral expansions derived for the real world for extrapolating quenched data to the chiral limit. The answer is, unfortunately, YES they should care.

The artifacts due to $\eta'$ loops can potentially invalidate all the extrapolations to the chiral limit. The hope is that since these are loop corrections and potentially large only in the limit $m_q \to 0$, therefore, there might exist a window in $m_q$ where the leading order chiral expansion is valid and sufficient, albeit with coefficients different from those in full QCD. Extrapolations of the quenched data from this range to the physical light $m_u$ may prove to be sensible, and the difference between the full and quenched coefficients taken as a measure of the goodness of the quenched approximation. With this goal in mind I analyze the existing quenched data in the range $m_s/4 - m_s$ and show that terms induced by the $\eta'$ are already visible and statistically significant.

In Section 9 I switch gears and review the status of $\overline{m}$ and $m_s$. The quenched Wilson fermion data for $\overline{m}$ is almost a factor of two larger, even at $\beta = 6.4$, than that for quenched staggered or $n_f = 2$ staggered or Wilson fermion data. The estimates of $m_s$ depend on whether $K$ or $K^*$ or $\phi$ is used to set the strange scale. These systematic differences are much larger than statistical errors and need to be brought under control.

## 2. QUENCHED CHIRAL PERTURBATION THEORY

Morel [5] gave a Lagrangian description of the quenched theory by introducing ghost quark fields with Bose statistics. This Lagrangian approach has been further developed by Bernard-Golterman into a calculational scheme. To the order we will be concerned with $\mathcal{L}_{BG}$ is

$$\begin{aligned}\mathcal{L}_{BG} &= \frac{f^2}{8}\mathrm{str}\left[\left(\partial_\mu\Sigma\partial_\mu\Sigma^\dagger\right) + 2\mu\left(M\Sigma + M\Sigma^\dagger\right)\right] \\ &+ \alpha_0\partial_\mu\Phi_0\partial_\mu\Phi_0 - m_0^2\Phi_0^2\end{aligned} \quad (1)$$

where $f = f_\pi = 131\ MeV$ is the pion decay constant, $\Sigma = \exp(2i\Pi/f)$, $M$ is the hermitian

Figure 1. The pseudoscalar propagator, (b) the hairpin vertex, and (c) the one bubble contribution to the $\eta'$ propagator in full QCD which after summation of all diagrams has the form shown.

$$\begin{array}{cc}
\rule{2cm}{0.4pt} & \dfrac{1}{p^2 + m^2} \\[1ex]
\longrightarrow\!\!\longleftarrow & \dfrac{1}{p^2 + m^2}\, m_0^2\, \dfrac{1}{p^2 + m^2} \\[1ex]
\longrightarrow\!\bigcirc\!\longleftarrow & \dfrac{1}{p^2 + m^2 + m_0^2}
\end{array}$$

quark mass matrix, $\mu$ sets the scale of the mass term, and str is the supertrace over quarks and ghost quarks. The last two terms involve the field $\Phi_0 = (\eta' - \tilde{\eta}')/\sqrt{2}$, where $\tilde{\eta}'$ is the ghost (commuting spin-1/2) field companion to the $\eta'$. These terms are treated as interactions and give rise to "hairpin" vertices (see Fig. 1) in the $\eta'$ propagator. This introduces two new parameters, $m_0^2$ and a momentum dependent coupling $\alpha_0 p^2$, in the quenched analysis. In the full theory this vertex and the tower generated by the insertion of bubble diagrams sum to give $\eta'$ its large mass, $m_0^2/(1-\alpha_0)$, while in the quenched theory the $\eta'$ remains a Goldstone boson and its propagator has a single and double pole.

The strength of the vertex, $m_0^2$, has been calculated on the lattice by the Tsukuba Collaboration [4] by taking the ratio of the disconnected to connected diagrams. It has also been determined using its relation to the topological susceptibility

$$m_0^2 = 2n_f \chi_t / f_\pi^2 = m_{\eta'}^2 + m_\eta^2 - 2m_K^2 \qquad (2)$$

measured on pure gauge configurations. These methods give $750 < m_0 < 1150\ MeV$. The parameter that occurs repeatedly in the chiral expansion of quenched quantities is $\delta \equiv m_0^2/24\pi^2 f_\pi^2$. Using $f_\pi = 131\ MeV$ and the above estimates for $m_0$ gives $0.14 \lesssim \delta \lesssim 0.33$. Current quenched data supports a value between $0.1 \lesssim \delta \lesssim 0.15$; different lattice observables give varying estimates due to statistical and systematic errors.

Let me first give an intuitive picture of why the $\eta'$ propagator gives extra contributions. The enhanced logs due to the $\eta'$ are infrared divergent, so it suffices to consider the $p^2 = 0$ limit in the $\eta'$ propagator. The single pole term is akin to the pion in the full theory, $1/m_\pi^2$, while the double pole term (due to the hairpin vertex diagram) is $\frac{1}{m_\pi^2} m_0^2 \frac{1}{m_\pi^2}$. Thus any time there is a normal correction term like $m_\pi^2 \mathrm{Ln} m_\pi^2$ from pion loops there will also be a singular term of the form $\frac{m_0^2}{m_\pi^2} m_\pi^2 \mathrm{Ln} m_\pi^2 = m_0^2 \mathrm{Ln} m_\pi^2 \sim \delta \mathrm{Ln} m_\pi^2$. This is exactly what one finds in the chiral expansion for $m_\pi^2$. Similarly, in the case of $m_{nucleon}$ the regular chiral correction is $\propto m_\pi^3$, and the $\eta'$ gives an extra term $\propto m_0^2 m_\pi$. My goal is to expose these extra terms in the present lattice data for different observables, and extract $\delta$ from them.

Further details on the formulation of the quenched chiral lagrangian and on the calculation of 1-loop corrections are given in Refs. [1] [3] [16]. The 1-loop corrections in the full and quenched theories show that
- the expansion coefficients are different,
- there are enhanced chiral logs,
- there are no kaon loops with strange sea quarks,
- values for parameters like $f$, $\mu$,.. are different

in the quenched expressions. I will assume that this difference is implicit in all subsequent discussion even when the same symbols are used for the two theories. Before addressing the consequences of these differences for the various physical quantities and their significance in the present data, I would like to mention the difference in the strategies, after 1-loop corrections have been calculated, of the two groups of alarmists. I find that knowing their respective emphasis helps in reading their papers.

Sharpe and collaborators focus on determining quantities that can be extracted reliably from quenched simulations. Using real world values to determine the chiral parameters (or commonly accepted ones if these are unknown parameters in $\chi$PT) they require that the chiral corrections are small in both the full and quenched expressions, as well as in their difference. Observables satisfying these conditions are the "good" candidates. Bernard and Golterman concentrate on testing quenched $\chi$PT by forming ratios of quantities which are (a) free of $O(p^4)$ terms in $\mathcal{L}_{cpt}$ and (b) independent of the ultraviolet cutoff used to regularize loop integration. The quenched chiral expansion of such ratios have terms proportional to the extra parameter $\delta$. Since these terms can be singular in the chiral limit, it is necessary to assume that there exists a window in $m_q$ where the 1-loop result is reliable. Then $\delta$ can be determined from fits to the quenched expression provided the fits to the quenched and full theory are significantly different.

## 3. $m_\pi^2$ VERSUS $m_q$

Gasser and Leutwyler [6] [11] show that in full QCD

$$m_\pi^2 = 2\mu m_q \big(1 + \frac{1}{2}L(m_\pi) - \frac{1}{6}L(m_\eta) + O(m_q)\big) \quad (3)$$

where $L(m) = m^2 \text{Ln}(m^2/\lambda^2)/8\pi^2 f^2$. Bernard and Golterman [1] and Sharpe [3] show that these logs are absent in the quenched approximation. Instead, for $\alpha_0 = 0$, they get

$$(m_\pi^2)_Q = 2\mu m_q \big(1 - \delta \text{Ln}(\frac{m_\pi^2}{\lambda^2}) + \ldots\big). \quad (4)$$

where $\lambda$ is some typical scale of $\chi SB$. This expression has been refined by Sharpe, who summed up the leading logs for the degenerate case $m_u = m_d = m_s$. We use his result [10]

$$\text{Ln}\frac{(m_\pi^2)_Q}{m_q} = c_0 - \frac{\delta}{1+\delta}\text{Ln} m_q + c_1 m_q + c_2 m_q^2 \quad (5)$$

to extract $\delta$ from a compendium of staggered fermion data at $\beta = 6.0$ [7][8][9]. Expressing all quantities in lattice units, the best fit gives

$$\text{Ln}\frac{(m_\pi^2)_Q}{m_q} = 1.54 - 0.044 \text{Ln} m_q + 1.2 m_q - 2.8 m_q^2. (6)$$

This implies that $\delta \approx 0.053$, i.e. much smaller than the value $\approx 0.3$ based on full QCD parameters; however, the breaking of flavor symmetry in staggered fermions has an interesting consequence for this analysis. The $\eta'$ operator is a singlet under staggered flavor, and different from the Goldstone pion which has flavor $\gamma_5$. Thus one should use the corresponding non-Goldstone pion mass in terms that come from the $\eta'$. In Fig. 2 the fit uses the $\tilde{\pi}$ (which has flavor $\gamma_4\gamma_5$) mass in the log term as it is better measured and consistent with the flavor singlet case. The result is

$$\text{Ln}\frac{(m_\pi^2)_Q}{m_q} = 1.35 - 0.13 \text{Ln} m_{\tilde{\pi}}^2 + 1.5 m_{\tilde{\pi}}^2 - 2.4 m_{\tilde{\pi}}^4. (7)$$

Figure 2. Fit to staggered $m_\pi^2/m_q$ data. At $\beta = 6.0$, an estimate of strange quark mass is $m_s a \approx 0.025$

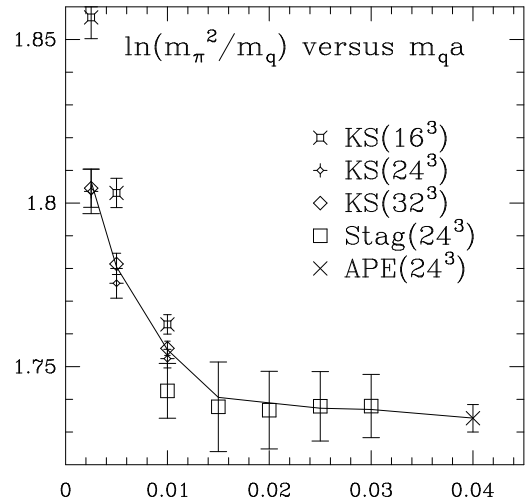

In this form the coefficient of the $\text{Ln} m_q$ term is $\delta$. Thus $\delta \approx 0.13$, a value consistent with the estimate 0.14 based on the calculation $m_0 = 750~MeV$. Also note that since the mass of the flavor singlet state, $\tilde{\pi}$, does not vanish as $m_q \to 0$, therefore, there is no singularity at finite $a$ due to the enhanced logs.

The above analysis shows that if one wanted to extract the value of $A_\pi$ in the expansion $m_\pi^2 = A_\pi m_q + \ldots$, then the quenched data would give a significantly different result depending on the kind of fit used. If one assumes that the 5 data points by the Staggered collaboration [7] represent a window in which $\chi$PT is valid and chiral corrections are negligible, i.e. the relation $m_\pi^2 = A_\pi m_q$ is sufficient (as expected at small enough $m_q$ in full QCD), then one gets $m_\pi^2 = 5.87 m_q$ [7], whereas Eq.7 gives $A_\pi \sim 3.9$, a significantly different value. The fit in Eq. 7 shows that over a range of $m_q$, the chiral log and higher order terms can conspire to produce a flat region.

Finite size effects in $m_\pi$ increase the value of $(m_\pi^2)_Q/m_q$, so one might attribute the 4% deviation at $m_q = 0.0025$ in Fig. 2 to this artifact. Fortunately, Kim and Sinclair [8] have obtained high statistics data for $m_q = 0.0025$, $0.005$, $0.01$ on lattices of size $L = 16$, 24, and 32 as shown in Fig. 2. There is clear indication of finite size effects on $L = 16$ lattices, but the near agreement between $L = 24$ and 32 data confirms that $L = 32$ is essentially infinite volume. To conclude, the data show that the lowest order chiral expansion has broken down and the effects of $\eta'$ logs are manifest for $m_q < m_s/2$. A similar analysis with Wilson fermions is not yet useful because the lowest $m_q$ used in simulations is $\sim 0.4 m_s$, i.e. the point where staggered fermions just start to show significant deviations.

## 4. BERNARD-GOLTERMAN RATIO $R$ AND $f_\pi$

The chiral behavior of $f_\pi$ in full QCD has been analyzed by Gasser and Leutwyler [11] to be

$$f_\pi = f\big[1 - L(m_\pi) - \frac{1}{2}L(m_K) + f(m_u + m_d + m_s)L_4 + m_u L_5\big] \quad (8)$$

where $L_4$ and $L_5$ are two $O(p^4)$ constants they introduce. In the quenched theory, with $\alpha_0 = 0$, Bernard-Golterman and Sharpe get

$$f_\pi = f\big(1 + m_u L_5\big). \quad (9)$$

The absence of pion and kaon chiral logs in the quenched expression is a $13 - 19\%$ effect (corresponding to the range $\lambda = 0.77 - 1$ GeV for the chiral symmetry breaking scale in $L(m)$) using full QCD parameters! To reliably compare full and quenched theories without ambiguities due to the cutoff $\Lambda$ and $O(p^4)$ terms, Bernard-Golterman construct, in a 4-flavor theory, the ratio

$$R \equiv \frac{f_{12}^2}{f_{11'} f_{22'}} \quad (10)$$

where $m_1 = m_{1'}$ and $m_2 = m_{2'}$. The $\chi$PT expression for $R$ in the full and quenched theories is

$$R^F = 1 + \frac{1}{32\pi^2 f^2}\left[ m_{11'}^2 \mathrm{Ln}\frac{m_{11'}^2}{m_{12}^2} + m_{22'}^2 \mathrm{Ln}\frac{m_{22'}^2}{m_{12}^2}\right]$$

$$R^Q = 1 + \delta \left[ \frac{m_{12}^2}{(m_{11'}^2 - m_{22'}^2)} \mathrm{Ln}\frac{m_{11'}^2}{m_{22'}^2} - 1\right]. \quad (11)$$

where the quantity within [ ] (called $X$) increases with the mass difference $m_2 - m_1$.

The quenched Wilson data for $R$ obtained by the LANL [12], UKQCD[13], and Bernard et al.[14] collaborations are shown in Figs. 3 and 4 versus the full and quenched expressions given in Eq.11. The slope of the fit to $R^Q$ gives $\delta$, while for $R^F$ the expected slope is unity. The data favor the quenched expression and give $\delta = 0.10(3)$.

Figure 3. The Bernard-Golterman ratio $R$ versus the full QCD expression given in Eq.11.

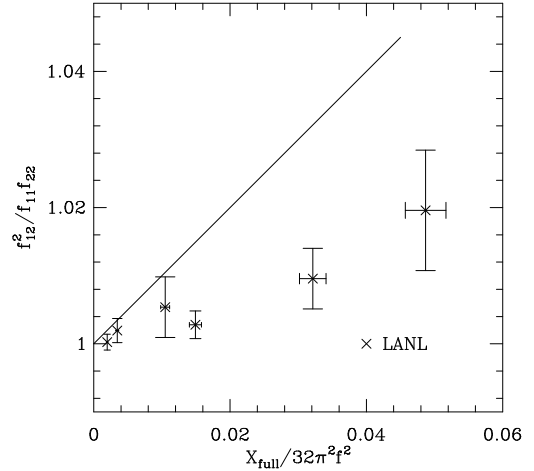

The caveat in this case is that the two points at largest $X^Q$ are obtained with $m_2 = 2m_s$, so one could argue that 1-loop $\chi$PT is not reliable for these masses. Barring this technicality, I believe that this quantity provides the cleanest determination of $\delta$.

## 5. BERNARD-GOLTERMAN RATIO $\chi$ AND $\langle \bar\psi\psi \rangle$

Bernard-Golterman construct a second quantity that is independent of $\Lambda$ and $O(p^4)$ terms

$$\chi = \frac{\langle \bar d d\rangle}{\langle \bar u u\rangle} - \left(\frac{M_{K^0}^2 - M_{K^+}^2}{M_{K^0}^2 - M_\pi^2}\right)\frac{\langle \bar s s\rangle}{\langle \bar u u\rangle} \quad (12)$$

for which $\chi$PT gives

$$\chi_{tree} = \frac{m_s - m_d}{m_s - m_u}, \quad (13)$$

$$\chi_Q = \chi_{tree} + \delta\big[\mathrm{Ln}\frac{m_u}{m_d} - \frac{m_d - m_u}{m_s - m_u}\mathrm{Ln}\frac{m_u}{m_s}\big],$$

$$\chi_F = \chi_{tree} + \frac{1}{8\pi^2 f^2}\big[M_{K^+}^2 \mathrm{Ln}\frac{M_{K^+}^2}{M_\pi^2} - \frac{m_s - m_d}{m_s - m_u}M_{K^0}^2 \mathrm{Ln}\frac{M_{K^0}^2}{M_\pi^2}\big].$$

Figure 4. The Bernard-Golterman ratio $R$ versus the quenched expression given in Eq.11.

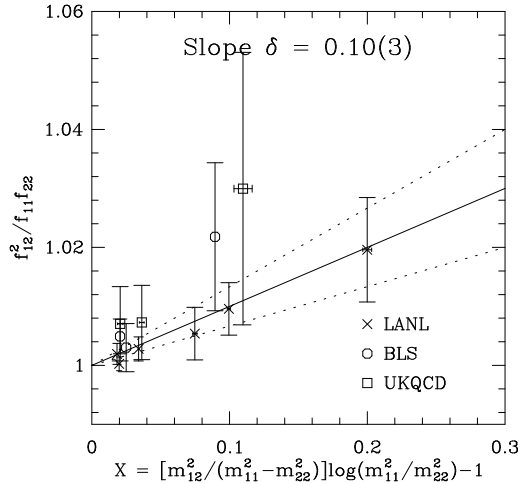

To evaluate these expressions requires data for the condensate at three values of $m_q$ and pseudoscalar masses for the combinations $\pi = uu$, $K^0 = sd$, $K^+ = su$. At present only the staggered [7] and Wilson [12] fermion simulations at $\beta = 6.0$ by the LANL collaboration have all the necessary data. Our results for $\delta = (\chi - \chi_{tree})/Y$, where $Y$ is the factor multiplying $\delta$ in the expression for $\chi_Q$ in Eq.14, are given in Table 1.

The staggered data have large errors and would give the wrong sign for $\delta$. (I have not taken into account the difference between Goldstone and non-Goldstone mass in terms that come from $\eta'$ loops.) With Wilson fermions the condensate in the chiral limit can be calculated in two ways, using the GMOR relation or the Ward Identity as explained in Ref. [15]. At finite $m_q$ there are lattice artifacts which we cannot control, nevertheless, the data give reasonable value for $\delta$. This is probably fortuitous and I believe that much better data is needed in order to extract $\delta$ from the chiral condensate.

Table 1
The Bernard-Golterman ratio X

|  | Staggered | Wilson(GMOR) | Wilson(WI) |
|---|---|---|---|
| $\chi$ | 0.549(30) | 0.608(6) | 0.614(5) |
| $\chi_{tree}$ | 0.517(14) | 0.620(2) | 0.620(2) |
| $\chi_F$ | 0.509(15) | 0.616(2) | 0.616(2) |
| $\delta$ |  | 0.10(5) | 0.05(4) |

## 6. CHIRAL EXTRAPOLATION OF THE NUCLEON MASS

The behavior of baryon masses has been calculated in $\chi$PT and has the general form [6]

$$M_B = \overline{M} + \sum c_i^{(2)} M_i^2 + \sum c_i^{(3)} M_i^3 \\ + O(m_\pi^4 \mathrm{Ln} m_\pi) \qquad (14)$$

where $M_i$ are $\pi$, $K$, $\eta$ meson masses. The term proportional to $M_i^3$ comes from pion loops and is $25\% - 50\%$ of $M_B$ for the octet. For example, using the results of Bernard et al. [19] one finds $M_N = 0.97 + 0.24 - 0.27$ respectively for the first three terms in Eq.14. Thus, the loop corrections in individual masses are large and one could question whether $\chi$PT is applicable at all to baryons. On the other hand $\chi$PT results for mass differences and the Gellmann-Okubo formula work very well, just as in the quark model. So, it is possible that the loop effects somehow conspire to just shift the overall scale, in which case $\chi$PT is useful and it is worthwhile examining the consequences of the quenched approximation.

Labrenz and Sharpe [16] have extended the Lagrangian approach of Bernard-Golterman to baryons using the "heavy-quark" formalism of Jenkins and Manohar [17]. They show that along with a modification of the $c_i$ in Eq.14 one gets a $m_0^2 m_\pi$ term due to $\eta'$ loops. The quenched expression for degenerate quark masses is (assuming $\alpha_0 = \gamma = 0$, where $\gamma$ is a parameter in the baryon sector of $\mathcal{L}_{cpt}$ and defined in [16])

$$M_B = \overline{M} + c^{(1)} \delta M_\pi + c^{(2)} M_\pi^2 + c^{(3)} M_\pi^3 + \ldots \quad (15)$$

where $c^{(1)} \sim -2.5$, $c^{(2)} \sim 3.4$, and $c^{(3)} \sim -1.5$ using full QCD values for the parameters.

Fits to lattice data using Eq. 15 are not very reliable because the number of light quark masses explored are typically $3-4$ and only the point at the heaviest mass (typically $m_q \gtrsim 2m_s$) shows any significant deviation from linearity. I find such 4-parameter fits to 4 points very unstable. For example, in the case of LANL data [12], even the different JK samples give completely different values of $c^{(i)}$. The best I could do was to fix one of the parameters and make a 3-parameter fit and then vary the fixed parameter to minimize $\chi^2$. The best fit (obtained by fixing any one of the less well determined coefficients, $c^{(1)}$, $c^{(2)}$ or $c^{(3)}$, as one gets the same final result on minimizing $\chi^2$) to the LANL data expressed in units of $GeV$ is shown in Fig.5 and gives

$$M_B = 1.16 - 0.36 M_\pi + 1.6 M_\pi^2 - 0.5 M_\pi^3. \qquad (16)$$

Figure 5. Fit to the LANL nucleon mass data.

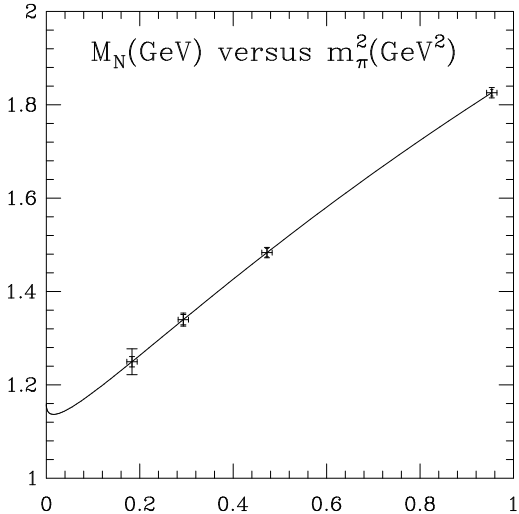

Assuming $c^{(1)} = -2.5$, Eq.16 gives $\delta \approx 0.14$. The same method applied to "012 sink" data from the GF11 collaboration[20] at $\beta = 5.93$ gives ( updated version of the fit presented in Ref.[16])

$$M_B = 1.18 - 1.0 M_\pi + 3.0 M_\pi^2 - 1.3 M_\pi^3 \qquad (17)$$

which implies that $\delta \approx 0.4$, and $c^{(2)}$ and $c^{(3)}$ have values close to those for full QCD.

## 7. THE KAON B PARAMETER

The kaon B parameter is a measure of the strong interaction corrections to the $K^0 - \bar{K}^0$ mixing. It is one of the best measured lattice quantities. For details of the phenomenology and of the lattice methodology I refer you to Refs.[21][22][23]. Here, I present a summary of just the chiral behavior.

Zhang and Sharpe [18] have calculated the chiral behavior of $B_K$ in both the full and quenched theories. The full QCD result is [23]

$$B_K = B\left[1-(3+\frac{\epsilon^2}{3})y\mathrm{Ln}y+by+c\epsilon^2 y+O(y^2)\right](18)$$

where $y = m_K^2/(8\pi^2 f^2) \approx 0.2$ and $\epsilon = (m_s - m_d)/(m_s + m_d)$ measures the degeneracy of $s$ and $d$ quarks. $B$ is the leading order value for $B_K$, which is an input parameter in $\chi$PT, and $b$ and $c$ are unknown constants. The quenched result [18]

$$\begin{aligned} B_K^Q &= B^Q\left[1 - (3+\epsilon^2)y\mathrm{Ln}y + b^Q y + c^Q \epsilon^2 y \right. \\ &\left. + \delta\left(\frac{2-\epsilon^2}{2\epsilon}\mathrm{Ln}\frac{1-\epsilon}{1+\epsilon} + 2\right)\right] \quad (19) \end{aligned}$$

has exactly the same form except for the additional term proportional to $\delta$, which is an artifact of quenching. The term proportional to $\delta$ is singular in the limit $\epsilon \to 1$, therefore extrapolations of quenched results to the physical non-degenerate case are not reliable. For $\epsilon = 0$ this term vanishes, so unless one works close to $\epsilon \to 1$ (for which there is little incentive in the quenched approximation), it is unlikely that we will, in the foreseeable future, be able to extract $\delta$ using Eq.19.

The constants $B$, $b$, $c$ are different in the full and quenched theories and cannot be fixed by $\chi$PT. Assuming $B = B^Q$, the coefficient of the chiral log term is the same for $\epsilon = 0$. This is the best agreement one can expect between the two theories. As a result Sharpe [23] advocates that $B_K$ with degenerate quarks is possibly a "good" quantity to calculate using the quenched theory, though systematic errors due to use of degenerate quarks are hard to estimate.

Using full QCD values, $3y\mathrm{Ln}y \approx 1$, so one can ask whether this normal chiral log is visible in the present data and whether it should be included in the extraction of $B_K$? With existing data it is hard to distinguish this term from the one linear in $y$ as the range of $m_K$ is not large enough to significantly affect the logarithm. Also, there exist data for $m_q \approx m_s/2$, so for degenerate quarks (which, as explained above, is the best one can do with the quenched theory) there is no need for an extrapolation.

Figure 6. Evidence of finite size effects in enhanced chiral logs in $B_V$.

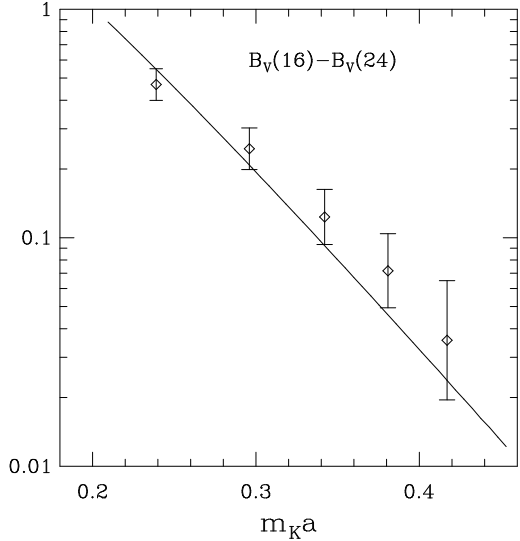

For staggered fermions $B_K$ can be written as the sum of two terms, $B_K = B_V + B_A$, each of which can be analyzed using $\chi$PT. These quantities are defined in Ref. [3] and are explicitly constructed such that they do not diverge as $1/m_K^2$ in the chiral limit. Both $B_V$ and $B_A$ have enhanced logs (terms proportional to Ln$y$ and not suppressed by powers of $y$) that have nothing to do with quenching, i.e. are not due to the $\eta'$. It is these logs, or more precisely the volume dependence of these logs, that has been seen in lattice data. Sharpe [3] has shown that this volume dependence is of the form

$$B_V(L) - B_V(\infty) = -(B_A(L) - B_A(\infty)) \quad (20)$$
$$\approx -b_2 \sqrt{\frac{2\pi}{m_K L}} \frac{6\mu^2 e^{-m_K L}}{8\pi^2 f^2}.$$

The constant $b_2$ is not well determined, but the shape of the $m_K$ dependence is. The staggered fermion data at $\beta = 6.0$ on $16^3$ and $24^3$ lattices [24] are shown in Fig. 6 and qualitatively confirm the expected finite size effects in the chiral logs.

## 8. MATRIX ELEMENT OF SINGLET AXIAL CURRENT IN THE PROTON

Ever since the measurement of the spin structure of protons using deep inelastic muon scattering from protons by the EMC collaboration[25], there has been much interest in the calculation of the forward matrix elements of the singlet axial current in the proton, $\langle \vec{p}, s | \bar{q} i \gamma_\mu \gamma_5 q | \vec{p}, s \rangle$. There are two possible Wick contractions that contribute to this matrix element ($ME$). These connected and disconnected diagrams are discussed in [27]. Since the disconnected diagram is hard to measure, Mandula [26] used the anomaly condition to derive the relation

$$\langle \vec{p}, s | A_\mu | \vec{p}, s \rangle s_\mu = N_f \frac{\alpha_s}{2\pi} \lim_{\vec{q} \to 0} \frac{-i|\vec{s}|}{\vec{q} \cdot \vec{s}} \times$$
$$\langle \vec{p}', s | \text{Tr} F_{\mu\nu} \tilde{F}_{\mu\nu}(\vec{q}) | \vec{p}, s \rangle \quad (21)$$

where $\vec{q} = \vec{p} - \vec{p}'$ and $s$ is the proton's spin vector. The hope was that it would be easier to measure the $ME$ of this purely gluonic operator. Since the $\eta'$ propagator contributes to this ME at tree level, the question arises whether Eq.21 is valid in the quenched approximation. The answer is NO [27]. Consider the Fourier transform of the anomaly relation

$$iq_\mu \langle \vec{p}', s | A_\mu(q) | \vec{p}, s \rangle = 2m_q \langle \vec{p}', s | P | \vec{p}, s \rangle +$$
$$N_f \frac{\alpha_s}{2\pi} \langle \vec{p}', s | \text{Tr} F\tilde{F} | \vec{p}, s \rangle. \quad (22)$$

Each of the three $ME$ in Eq.22 can be parameterized in terms of form-factors as

$$\langle \vec{p}', s | A_\mu(q) | \vec{p}, s \rangle = \bar{u} i \gamma_\mu \gamma_5 u G_1^A - i q_\mu \bar{u} \gamma_5 u G_2^A,$$
$$\langle \vec{p}', s | P | \vec{p}, s \rangle = \bar{u} \gamma_5 u G^P,$$
$$\langle \vec{p}', s | \text{Tr} F\tilde{F} | \vec{p}, s \rangle = \bar{u} \gamma_5 u G^F. \quad (23)$$

In the quenched approximation the singularities in these form factors for on-shell $ME$ with respect to $q^2$ and due to the $\eta'$ propagators are

$$G_1^A(q^2) \quad \text{no } \eta' \text{ poles},$$
$$G_2^A(q^2) = \frac{a_2}{(q^2 - m_{\eta'}^2)^2} + \frac{a_1}{(q^2 - m_{\eta'}^2)} + \tilde{G}_2,$$
$$G^P(q^2) = \frac{p_2}{(q^2 - m_{\eta'}^2)^2} + \frac{p_1}{(q^2 - m_{\eta'}^2)} + \tilde{P},$$
$$G^F(q^2) = \frac{f_1}{(q^2 - m_{\eta'}^2)} + \tilde{F}. \quad (24)$$

Equating the single and double pole terms gives two relations. Using these and taking the double limit, $q^2 \to 0$ and $m_q \to 0$, gives

$$2M_P G_1^A(q^2 = 0) = -a_1 + N_f \frac{\alpha_s}{2\pi}\tilde{F} \qquad (25)$$
$$= \frac{2m_q}{m_{\eta'}^2}\left(\frac{p_2}{m_{\eta'}^2} - p_1\right) + N_f \frac{\alpha_s}{2\pi}\left(\tilde{F} - \frac{f_1}{m_{\eta'}^2}\right).$$

The term proportional to $N_f \alpha_s/2\pi$ diverges in the chiral limit and there is no obvious way of extracting the physical answer from it alone. Thus the method fails in the quenched theory.

In the full theory, there are no double poles and an analogous analysis gives

$$2M_P G_1^A(q^2 = 0) = -a_1 + N_f \frac{\alpha_s}{2\pi}\tilde{F}$$
$$= N_f \frac{\alpha_s}{2\pi}\left(\tilde{F} - \frac{f_1}{m_{\eta'}^2}\right), \qquad (26)$$

which justifies the use of the anomaly relation.

## 9. MASSES OF LIGHT QUARKS

In order to extract light quark masses from lattice simulations we use an ansatz for the chiral behavior of hadron masses. Theoretically, the best defined procedure is $\chi$PT which relates the masses of pseudoscalar mesons to $m_u$, $m_d$, $m_s$.

The overall scale $\mu$ in the mass term of Eq.1 implies that only ratios of quark masses can be determined using $\chi$PT. The predictions from $\chi$PT for the two independent ratios are [6] [32]

|  | Lowest order | Next order |
|---|---|---|
| $(m_u + m_d)/2m_s$ | $\frac{1}{25}$ | $\frac{1}{31}$ |
| $(m_d - m_u)/m_s$ | $\frac{1}{44}$ | $\frac{1}{29}$ |

In Lattice QCD it is traditional to make fits to the pseudoscalar spectrum assuming $m_{12}^2 = A_\pi(m_1 + m_2)$ and using either $m_\rho$ or $f_\pi$ to set the scale. (The expression in Eq.5 is not relevant for this discussion since most quenched simulations have $m_q \geq m_s/2$.) A consequence of using just the linear term is that the ratio $m_s/\overline{m} = 25$, i.e. these fits can be used to extract either $\overline{m} = (m_u + m_d)/2$ or $m_s$ by using the physical masses for $m_\pi$ or $m_K$, but not both. (One would get a different number if $O(m_q^2)$ and chiral log terms are included in the relation.) Furthermore, since lattice calculations are done in the isospin limit, $m_u = m_d$, therefore $\chi$PT can be used to predict only one quark mass. The mass I prefer to extract, barring the complications of quenched $\chi$PT, is $\overline{m}$ as it avoids the question whether lowest order $\chi$PT is valid up to $m_s$. Akira Ukawa reviewed the status of $\overline{m}$ at LATTICE92 [28] and I present an update on it.

To convert lattice results to the continuum $\overline{MS}$ scheme I use

$$m_{cont}(q^*) = m_{latt}(a)\left[1 - \frac{g^2}{2\pi^2}\big(log(q^*a) - C_m\big)\right] (27)$$

where the renormalization scale $\mu$ is the same as $q^*$ (defined in [29]) and chosen to be $\pi/a$, $C_m = 2.159$ for Wilson [30] and 6.536 for staggered fermions [31], and the rho mass is used to set the scale. (I have not used the tadpole improvement factor of $U_0$ [29] in $C_m$ and $m_{latt}$ as this factor cancels in perturbation theory and is a small effect otherwise.) The value of boosted $g^2$ I use in Eq.27 is [29]

$$\frac{1}{g^2} = \frac{\langle plaq \rangle}{g_{latt}^2} + 0.025 \qquad (28)$$

which is consistent with the continuum $\overline{MS}$ scheme value at $Q = \pi/a$

$$\frac{g^2(Q)}{16\pi^2} = \frac{1}{\beta_0 \text{Ln}(\frac{Q^2}{\Lambda^2})}\left(1 - \frac{\beta_1 \text{Ln}[\text{Ln}(\frac{Q^2}{\Lambda^2})]}{\beta_0^2 \text{Ln}(\frac{Q^2}{\Lambda^2})}\right) \qquad (29)$$

provided I use $\Lambda = 245~MeV$ and $190 MeV$ for $n_f = 0$ and 2 theories respectively. Note that the choice of $\Lambda$, $q^*$, and the constant 0.025 in Eq.28 are interrelated and, at this order, one can trade changes between them. Finally, all the results are run down to $Q = 2~GeV$ using

$$\frac{m(Q)}{m(q^*)} = \left(\frac{g^2(Q)}{g^2(q^*)}\right)^{\frac{\gamma_0}{2\beta_0}} \times$$
$$\left(1 + \frac{g^2(Q) - g^2(q^*)}{16\pi^2}\Big(\frac{\gamma_1\beta_0 - \gamma_0\beta_1}{2\beta_0^2}\Big)\right). (30)$$

The status of calculations of $\overline{m}(2~GeV)$ for quenched Wilson [35] [20] [36] [37] [12] [38], quenched staggered [39] [9] [40] [7] [8], dynamical ($n_f = 2$) Wilson [41] [42], and dynamical ($n_f = 2$) staggered fermions [40] [43] [44] is shown in Fig. 7. I have suppressed error bars as I want to first emphasize key qualitative features. The quenched staggered and the $n_f = 2$ Wilson and staggered give $\overline{m} = 2 - 3\ MeV$ and are roughly consistent; however, the quenched Wilson results seem to approach that value from above and even at $\beta = 6.4$ are significantly higher. (The recent result $m_s = 128(18)\ MeV$ by Allton et al. [45] for both Wilson and Sheikholeslami-Wohlert actions at $\beta = 6.0$ and $6.2$ is consistent with the results in Fig. 7 once one notes that $\overline{m} = m_s/25$.) I believe that, at this stage, it is important to understand why the quenched results with the Wilson action are so different from the rest!

Figure 7. $\overline{m}$ extracted using $m_\pi$ data with the scale set by $m_\rho$.

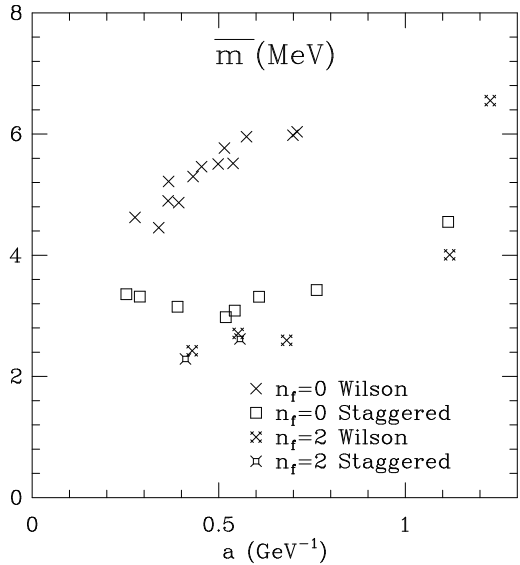

An alternative to using $m_q = Z_{mass} m_q^L = Z_s^{-1} m_q^L$ to calculate the quark masses with Wilson fermions is to use the Ward identity [34][15]

$$m_q = \frac{Z_A}{Z_P} \frac{m_\pi}{2} \frac{\langle A_4(\tau) P(0) \rangle}{\langle P(\tau) P(0) \rangle}. \qquad (31)$$

Using the perturbative values for $Z_A$ and $Z_P$ (with $q^* = \pi/a$ and boosted $g^2$ defined in Eq.28) the LANL Wilson data [12] gives $\overline{m} = 3.53(10)\ MeV$ in contrast to $\overline{m} = 5.15(15)\ MeV$ shown in Fig. 7. The statistical errors are calculated using a single elimination JK with a sample of 100 lattices of size $32^3 \times 64$, so the difference is significant. The Rome collaboration [45] has found a similar discrepancy and argue that it can be resolved if one uses the non-perturbative value for $Z_P$, which they advocate calculating using matrix elements of the operators between quark states in a fixed (Landau) gauge. Their results indicate that perturbation theory (including tadpole improvement) fails for $Z_P$. The two methods for extracting $\overline{m}$ give consistent results once the non-perturbative value of $Z_P$ is used.

Having fixed $\overline{m}$ one can extract $m_s$, $m_c$, and $m_b$ using, for example, $K^*$, $D$, and $B$ meson masses provided it is assumed that these masses are linear in the light quark mass and in the heavier quark mass around the physical value. Alternately, one can use $m_\phi$, $J/\psi$, and $\Upsilon$ spectrum to get these quark masses directly without needing to extrapolate in the light quark mass. The results for $m_c$ and $m_b$ have been reviewed by Sloan [33] at this conference so I will only analyze the data for $m_s$ and compare these estimates to $25\overline{m}$. Note that the same data used to compile Fig. 7 is used to calculate $m_s$ from $m_{K^*}$ and $m_\phi$. The procedure for translating the value to $2\ GeV$ in the $\overline{MS}$ scheme is also the same. The results in Fig. 8 show that that the estimate of $m_s$ from $m_\phi$ is systematically higher by $15-20\%$ compared to $25\overline{m}$.

I will use the LANL data[12] to show that the systematic errors due to choice of hadron used to set the scale of the strange quark are now a dominant source of error. We find that, in the $\overline{MS}$ scheme at $2\ GeV$, $m_s = 25\overline{m} = 129(4)\ MeV$ using $M_K$, $m_s = 151(15)\ MeV$ using $M_{K^*}$, and $m_s = 157(13)\ MeV$ using $M_\phi$. Note that the latter two estimates give $m_s/\overline{m} \sim 30$, which is much closer to the "Next Order" prediction of $\chi PT$. The larger errors in these cases reflect the fact that on the lattice masses of pseudoscalar mesons are measured with much better statistical accuracy than those of vector mesons.

## 10. CONCLUSIONS AND COMING ATTRACTIONS

The analysis of various quenched quantities show that the parameter $\delta$ characterizing the hairpin vertex in the $\eta'$ propagator lies in the range $0.1 - 0.2$. As a result, for $m_q \lesssim m_s/2$ I find significant deviations from the lowest order chiral behavior in $m_\pi^2/m_q$. Therefore, I conclude that extrapolation of quenched data, obtained with $m_q \leq m_s/2$, to the chiral limit cannot be done simply using full QCD formulae for quantities which have large contributions from enhanced logs. For quantities like the matrix element of the singlet axial vector current using the Adler-Bell-Jackiw anomaly, the quenched approximation fails altogether.

Figure 8. Comparison of $m_s$ extracted using $m_\phi$ and $m_s = 25\overline{m}$. The data are for quenched Wilson simulations.

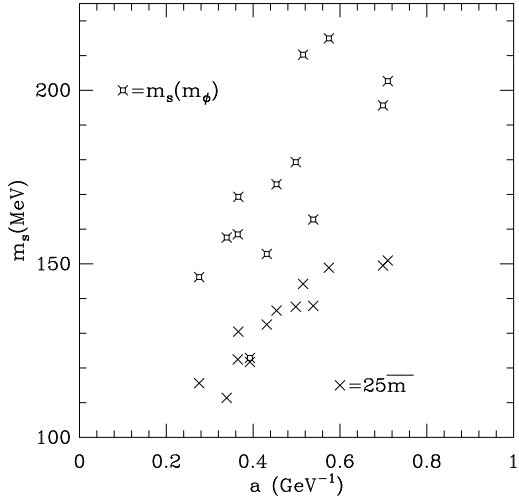

The alarmists are busy calculating 1-loop corrections to other quantities to determine what can be extracted reliably from quenched simulations. Bernard and Golterman have extended the results presented at LATTICE93 [46] and calculated chiral corrections to the energy of two pions in a finite box as derived by Lüscher [47]. They find terms at $O(1)$ and $O(1/L^2)$, whose contribution could be substantial, in addition to modifications of the $O(1/L^3)$ term which is related to the $\pi - \pi$ scattering amplitude [48]. Sharpe and Labrenz have extended the analysis of baryons to include the $\Delta$ decuplet [49]. Booth [50] and Zhang and Sharpe [18] have calculated corrections to heavy-light meson properties like $f_B$ and $B_B$. These new results and more data should provide a clearer picture of what is possible with quenched QCD by LATTICE 95.

In the calculations of light quark masses we need to understand the factor of two difference between the quenched Wilson and staggered data. On the other hand, the quenched staggered data is consistent with the $n_f = 2$ Wilson and staggered data. The analysis presented here leaves open the question — is the agreement between quenched Wilson (and $O(a)$ improved SW action) data with the phenomenologically favored estimates of $\overline{m}$ (or equivalently $m_s$) fortuitous and an artifact of strong coupling? If so, then the $n_f = 0, 2$ staggered and $n_f = 2$ Wilson data give an estimate of $\overline{m}$ that is $2 - 3$ times smaller than the commonly accepted phenomenological value.

The systematic errors due to the choice of hadron mass used in determining $m_s$ are significant. Using $m_{K^*}$ or $m_\phi$ to extract $m_s$ gives a $\sim 20\%$ larger value than that obtained from $m_K$. Even though the statistical errors are larger when extracting $m_s$ from vector mesons, these estimates provide information beyond the lowest order $\chi$PT result $m_s = 25\overline{m}$. Phenomenological estimates involving extrapolation to strange quark mass need to take this systematic difference into account.

## 11. ACKNOWLEDGEMENTS

I thank Claude Bernard, Tanmoy Bhattacharya, Maarten Golterman, and especially Steve Sharpe for many enlightening discussions. I thank Martin Lüscher for reminding me that the $\eta'$ is a staggered flavor singlet. I am grateful to Claude Bernard, Richard Kenway and Don Sinclair for providing unpublished data and to Akira Ukawa for the quark mass data he presented at LATTICE92. The staggered and Wilson fermion calculations by the LANL group have been done as part of the DOE HPCC Grand Challenges program. The recent Wilson fermion simulations on $32^3 \times 64$ lattices have been done on the CM5 and we gratefully acknowledge the tremendous support provided by the ACL at Los Alamos and by NCSA at Urbana-Champaign.